# Title: GHz spiking neuromorphic photonic chip with *in-situ* training

**Authors:** Jinlong Xiang[1]†, Xinyuan Fang[2]†, Jie Xiao[1], Youlve Chen[1], An He[1], Yaotian Zhao[1], Zhenyu Zhao[1], Yikai Su[1], Min Gu[2*], Xuhan Guo[1*]

**Affiliations:**

[1]State Key Laboratory of Photonics and Communications, School of Information and Electronic Engineering, Shanghai Jiao Tong University; Shanghai, 200240, China.

[2]School of Artificial Intelligence Science and Technology, University of Shanghai for Science and Technology; Shanghai, 200093, China.

*Corresponding author. Email: gumin@usst.edu.cn; guoxuhan@sjtu.edu.cn;

†These authors contributed equally to this work.

## Abstract

Neuromorphic photonic computing represents a paradigm shift for next-generation machine intelligence, yet critical gaps persist in emulating the brain's event-driven, asynchronous dynamics—a fundamental barrier to unlocking its full potential. Here, we report a milestone advancement of a photonic spiking neural network (PSNN) chip, the first to achieve full-stack brain-inspired computing on a complementary metal oxide semiconductor-compatible silicon platform. The PSNN features transformative innovations of gigahertz-scale nonlinear spiking dynamics, *in-situ* learning capacity with supervised synaptic plasticity, and informative event representations with retina-inspired spike encoding, resolving the long-standing challenges in spatiotemporal data integration and energy-efficient dynamic processing. By leveraging its frame-free, event-driven working manner, the neuromorphic optoelectronic system achieves 80% accuracy on the KTH video recognition dataset while operating at ~100x faster processing speeds than conventional frame-based approaches. This work represents a leap for neuromorphic computing in a scalable photonic platform with low latency and high throughput, paving the way for advanced applications in real-time dynamic vision processing and adaptive decision-making, such as autonomous vehicles and robotic navigation.

## Introduction

Inspired by the insights from neuroscience, neuromorphic computing emulates neural networks at the hardware level rather than simulating them on digital computers. This approach holds significant promise in overcoming the fundamental bottlenecks of the von Neumann architecture and addressing the ever-growing demand for computational power. Neuromorphic electronic platforms, such as the classic TrueNorth[1], and the most recent Tianjic[2], have achieved impressive energy-efficiency enhancement via spike-based information processing, compared to traditional digital processors. However, as Moore's law approaches its end and Dennard scaling concludes[3], electronic links face inherent limitations in speed, bandwidth, and crosstalk.

Photonics can overcome the limitations of electronics with intrinsic parallelism, ultra-wide bandwidth, and near-zero latency[4,5], making possible fascinating photonic computing systems for image classification[6,7], video recognition[8-11], and artificial content generation[12]. Despite recent exciting advances, those computer-science-oriented photonic approaches merely emulate the connectivity and functionalities of biological systems, while failing to replicate the asynchronous, event-driven, and dynamic processing capacities of human brains. In contrast, the neuroscience-oriented photonic spiking neural network (PSNN) works in a more biologically plausible way with collocated memory and computation, event-based information coding and processing, and synaptic plasticity learning[13]. PSNNs are particularly well-suited for processing dynamic spatiotemporal information in an efficient spike-driven fashion, making them ideal for time-varying application scenarios, although their development remains in its early stages.

To date, various time dynamics such as photocarriers, optical modes, and polarization competition in semiconductor lasers have been explored to mimic the biological behaviors of spiking neurons[14-21]. While the operation speed can reach a gigahertz (GHz) level, the network scale has been hindered due to the poor compatibility of III-V materials with mutual microelectronic fabrications. Silicon photonics, compatible with the complementary metal-oxide semiconductor (CMOS) technology, holds great potential to build neuromorphic computing systems on a large scale far beyond what can be attainable on fiber, III-V, or holographic platforms, and has witnessed great achievements in photonic deep neural networks[22-24] and tensor accelerators[25-27]. Among the available solutions, phase-change materials have offered a compact way to integrate both nonlinear activation and memory units into neuro-synaptic systems with self-learning capacities and structure adaptivity[28,29]. Additionally, all-optical[30,31] and optoelectronic[32,33] silicon microrings have been demonstrated to exhibit excitable properties of spiking neurons. However, it still remains tremendously challenging to implement GHz-level spiking information processing with silicon photonic integrated circuits, owing to the lack of ultrafast, scalable, and integrable spiking neurons.

Here we present the first, to the best of our knowledge, GHz spiking neuromorphic photonic chip on a silicon platform that operates in a full-stack bioinspired manner. The PSNN chip can perform ultrafast static image classification and frame-free dynamic video recognition through the following three key innovations. For the fundamental computing primitive, we have designed a scalable optoelectronic microring spiking neuron with a maximum firing rate of 4 GHz, which was previously only attainable with III-V lasers. For the network learning, we have realized an experimental scheme for in-situ training of PSNNs with supervised bio-plausible synaptic plasticity, by incorporating a closed communication and control loop. For the information representation, we have developed a retina-inspired spike encoding technique to efficiently convert video frames into sparse and informative spike trains, significantly reducing the information redundancy and naturally aligning with the event-driven nature of PSNNs. With these innovations combined, we have advanced toward challenging real-time video recognition, achieving 80% accuracy on the KTH benchmark dataset. Notably, the processing speed of frames-per-second (FPS) has remarkably improved by approximately two orders of magnitude compared to prior frame-based computing paradigms. The presented neuromorphic photonic system features high integration, low power consumption, and excellent scalability on the CMOS-compatible silicon platform with a brain-like neurosynaptic computing framework. Our work marks a milestone achievement in the development of neuromorphic photonic computing and will surely open up new opportunities for intelligent photonic applications.

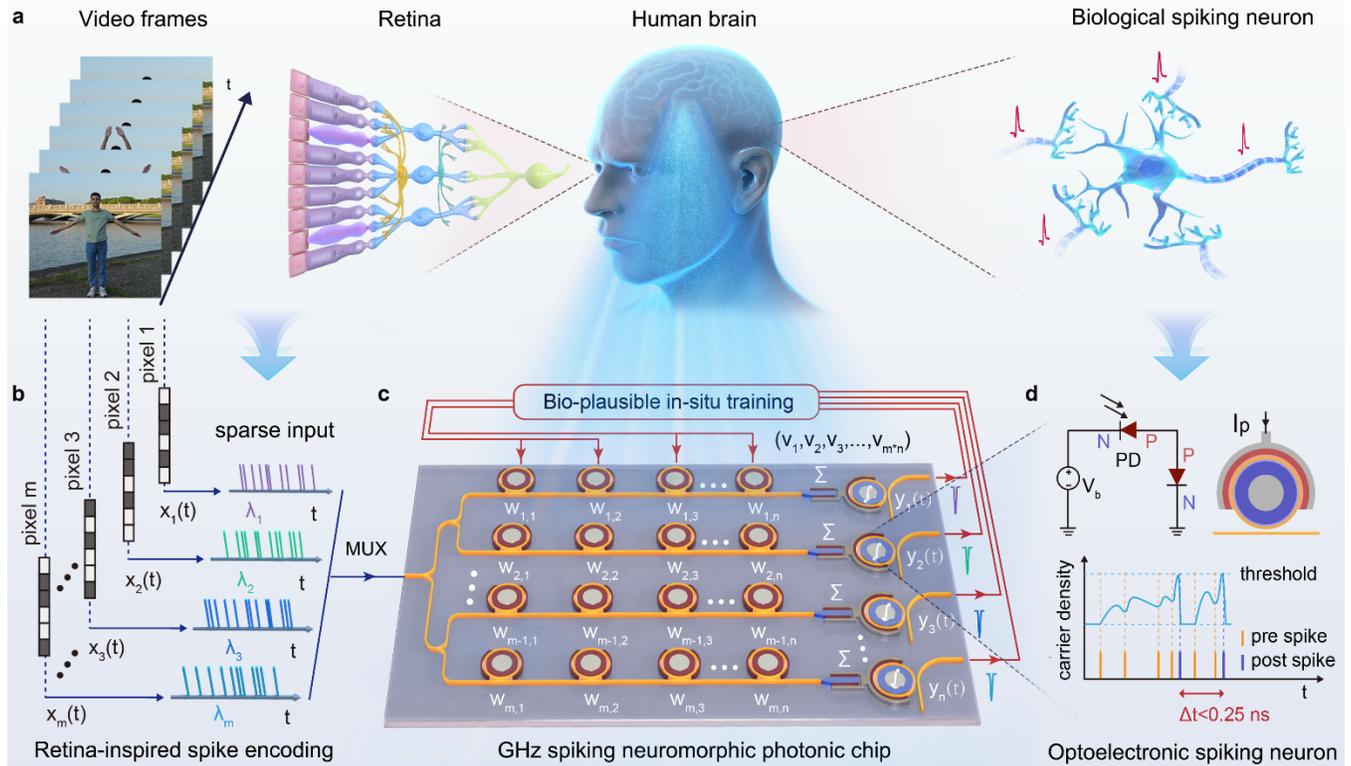

**Fig. 1 | GHz spiking neuromorphic photonic chip for real-time video processing. a,** Mechanisms of the human visual system. **b,** Retina-inspired spike encoding method that only detects feature changes. **c,** The photonic spiking neural network (PSNN) chip is implemented with a wavelength-multiplexed neurosynaptic framework and can be in-situ trained with supervised synaptic plasticity rules. **d,** The electrically-driven microring can accumulate free carriers injected from the photodetector, and emit optical spike responses at a GHz-scale speed once the carrier density crosses its excitability threshold.

## Results

### Silicon neuromorphic photonic chip

Figure 1a schematically shows the mechanisms of the human visual system. The retina is composed of an outer nuclear layer (ONL, photoreceptors), an inner nuclear layer (INL, on/off bipolar cells), and a ganglion cell layer (GCL). Dynamic input streams from the ever-changing real world are first sensed by the ONL, then go through pre-processing such as feature extraction and spike encoding in the INL and GCL. Eventually, the pre-processed information is transmitted to the human brain for high-level processing like decision-making. As the most important computational primitive for information transfer or exchange, the spiking neuron can accumulate stimuli from multiple synapses and generate spike activations once its membrane potential reaches the threshold,

Analogous to the principles of biological systems, the spiking neuromorphic photonic chip works in a fully bio-plausible manner. Input videos after pre-processing are transformed into spatiotemporal spike trains with a retina-inspired encoding technique in Fig. 1b that only detects feature changes (see the last section for details). Then, time-varying pixel values are modulated onto wavelength-multiplexed optical carriers, distinguished with different colors, and coupled into the PSNN chip for spiking information processing. Specifically, the PSNN chip in Fig. 1c is organized with a scalable broadcast-and-weight computing framework[34]. Following the feed-forward architecture, the multiplexed input is first broadcast to all POST neurons in the output layer by power splitters, then weighted in parallel with microring resonator (MRR) synapse arrays, and finally detected by photodetectors. Especially, the timing information of pre- and post-spikes can be extracted in the closed feedback loop and applied to in-situ training with a supervised synaptic plasticity rule. As diagramed in Fig. 1d, the generated photocurrent $I_p$, representing the summed weighted signals, modulates the microring modulator (MRM) transmission via free carrier injection. If the gradually accumulated carrier density in the MRM crosses a certain threshold, the electrically driven MRM will generate neuron-like spiking responses. After an excitation, the carrier density will rapidly recover to the equilibrium level and then go through the accumulation process again. Optical spikes with sub-ns intervals can be activated faithfully, thus offering the essential ultrafast nonlinear activation in PSNNs. Note that the optical-electrical–optical conversion allows flexible bias settings for each MRM neuron, and also mitigates the problem of insertion losses in optical links. The PSNN is fully compatible with mainstream silicon platforms, hence providing critical aspects of feasibility and economies of scale for compelling neuromorphic optoelectronic applications.

**Experimental characterization**

The designed neuromorphic photonic chip is fabricated with a commercial silicon photonics process in CUMEC within a multi-project wafer (https://service.cumec.cn/soi_csip130c). Figure 2a presents the packaged PSNN chip, where a fiber array and three FPC connectors are used for optical and electrical inputs/outputs, respectively. As shown in Fig. 2b, the PSNN chip comprises two photonic neuro-synaptic networks, each consisting of 10 MRM spiking neurons and 10 MRR weight banks.

The connection strength of a silicon MRR synapse is represented by its optical transmission at the through port and can be effectively tuned with in-ring n-doped photoconductive heaters. Given that MRRs are sensitive to fabrication imperfections and thermal crosstalk, a two-step calibration method is utilized to precisely control the multi-channel MRR weight bank[35] (See Supplementary Note 2 for more details). To benchmark the weight loading accuracy, we generate 160 random weight matrices distributed in the 0~1 interval and acquire 50 groups of actual weight values within 20 minutes. The bit precisions for all 10 channels are higher than 8 bits (Fig. 2d), which is comparable to that used in DSP ASICs. Figure 2e presents the measured weights as a function of target weights for the first channel, with all points tightly concentrated along the diagonal line. The attached histograms show the distribution of errors between measured and desired weight values, where a standard deviation of 0.00335 indicates negligible inter-channel crosstalk after calibration. It should be noted that the weight range can be scaled to [-1, 1] by forming add-drop MRRs and balanced photodetectors into a complementary configuration[36].

Figure 2f shows the microscope photo of a fabricated MRM neuron, which can integrate the current from a photodetector and fire optical spikes once its carrier density exceeds the excitability threshold. The temporal characteristics can be described by an electro-optical co-simulation model[33], where free carriers are employed to bridge the electrical circuit model of a p-n diode and the optical time-domain model of a passive MRR (See details in Supplementary Note 4). To explore the excitable mechanism of an electrically-driven MRM, bifurcation analysis is performed on the simplified dimensionless rate equations. As shown in Fig. 2g, N and $N_{pn}$ represent total free carriers in the MRM and injected carriers from the p-n diode, respectively. It can be seen that a subcritical Hopf bifurcation point is identified at $N_{pn}=2.53\times10^{22}$ m$^{-3}$ with a stable fixed point centered in a stable limit cycle, which implies that the carrier-injection MRM can emulate a Class II resonate-and-fire spiking neuron[31]. Importantly, no Hopf bifurcation point is observed in the region of $N_{pn}<0$, indicating carrier-depletion MRMs are unable to mimic excitable spiking neurons. Similar to III-V laser neurons, the output characteristics of electrically-driven MRM neurons are dominated by fast carrier dynamics, and their operation speed is largely constrained by the carrier recombination process. As the carrier lifetime in silicon waveguides can be effectively tailored with ion implantation, we have increased the doping concentration from the level of $10^{17}$ cm$^{-3}$ in prior work[33] to $10^{18}$ cm$^{-3}$ and optimized the electro-optical RC bandwidth by dedicatedly engineering the doping profile, leading to a significantly improved firing rate of 4 GHz.

When working near its excitable regime, the optoelectronic MRM neuron can exhibit typical neuron-like behaviors. We present reference electrical signals and measured output waveforms with red and blue lines, respectively. As shown in Fig. 2h, the MRM generates a strong spike response to the first perturbation pulse but remains silent for the second stimulus, which reveals the essential spiking property of the excitability threshold. Besides, the MRM neuron can be successfully excited by three closely-spaced stimuli or a single sub-threshold stimulus with a longer time duration, which proves the MRM can accumulate the perturbation energy of external stimuli within a small time window, i.e., the temporal integration property. Moreover, similar spike responses can be excited by a pair of perturbation pulses with a time interval of 0.25 ns, as depicted in Fig. 2i. Therefore, the operation speed of our MRM neuron reaches up to 4 GHz, which is even superior to typical values of 1~3 GHz for III-V laser neurons[19,20,37]. Note that stable and reproducible spike responses can be obtained for the same perturbation stimuli (See Supplementary Note 7 for more details), which lays a solid foundation for complex spiking information processing. Figure 2j compares state-of-the-art photonic spiking neurons, where the integration scale is divided into three levels. The MRM neuron with impressive GHz speed and good compatibility to commercial foundry process has constituted the final building block to construct large-scale PSNN chips for fantastic neuromorphic photonic applications.

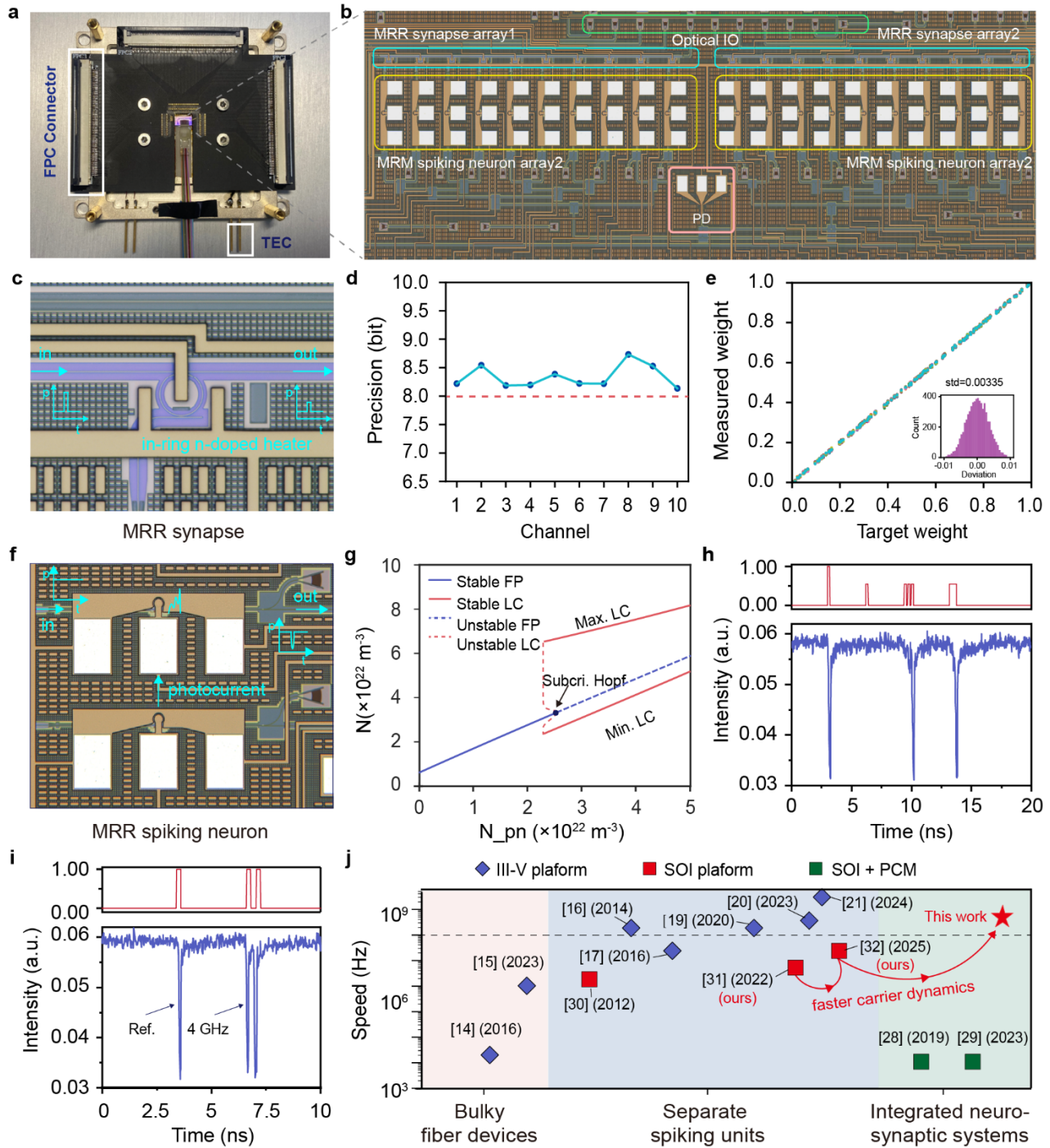

**Fig. 2 | Characterization of the neuromorphic photonic chip. a**, Packaged PSNN chip, with FPC connectors for electronic links, a fiber array for optical inputs/outputs, and a TEC for temperature stabilization. **b**, Microscope photo of the PSNN chip, consisting of two MRM neuron arrays and two MRR synapse arrays. **c**, Zoom-in view of a silicon MRR synapse. **d**, Weight-loading precisions of 10 wavelength channels after calibration. **e**, Measured weights of the first channel as a function of randomly

generated target weights, with the distribution of errors shown in the insert. **f**, Zoom-in view of two MRM neurons. **g**, Bifurcation analysis of the electrically-driven MRM, where stable and unstable fixed points (FPs) or limit cycles (LCs) are indicated with solid or dashed lines, respectively. Experimental results of **h**, excitability threshold & temporal integration, and **i**, refractory period. **j**, Performance comparison between existing photonic spiking neurons, where the integration scale is divided into three levels: bulky fiber devices[14,15], separate spiking units[16-21,30-32], and integrated neuro-synaptic systems[28,29].

## Bio-plausible *in-situ* training

Implementing neuromorphic algorithms solely with photonics makes possible high-performance signal processing and ultrafast interference at near-terahertz rates. A hardware-algorithm collaborative computing scheme has been reported for spiking pattern recognition[20], but the weights are trained offline and mapped to optical devices. Owing to the inherent systematic errors in analog computing from various sources, the in silico-trained PSNN model cannot faithfully represent the physical system, leading to significant performance degeneration during direct deployment. To overcome the simulation-reality gap in offline training, in-situ approaches that train PSNNs with experimental feedback are much preferred for practical applications.

We develop a supervised in-situ training paradigm for PSNNs, which incorporates the input spike encoding, on-chip spiking processing, output acquisition, and post-processing, as well as the final weight updating into a closed communication and control pipeline, as schematically shown in Fig. 3a. All instruments, i.e., tunable lasers, the arbitrary waveform generator (AWG), the multi-channel voltage source (MVS), the power meter, and the digital signal oscilloscope (DSO), are programmed to communicate and exchange information with the host PC via the PyVISA package[38]. During each iteration, the Python program first transforms input patterns into AWG spike trains $I_m(t)$ and modulates them on multi-wavelength carriers. The multiplexed optical signals go through on-chip weighting by MRR synapses and nonlinear spike activation by MRM neurons, and then are converted back to electrical signals by photodetectors; After that, the main program fetches captured waveforms from DSO and performs necessary data post-processing; The activation state of each POST neuron $O_n(t)$ is determined by comparing with a predefined excitability threshold, and precise timing relations between pre- and post-synaptic spikes can be obtained after synchronization; Finally, the weight matrix is updated from $\omega_j$ to $\omega_j + \eta \cdot \Delta \omega_j$ based on the modified remote supervised method (ReSuMe) learning rule, and thermal-tuning voltages $V_j$ applied to MRR synapses are modified via the MVS accordingly.

To verify our in-situ training scheme, a small PSNN composed of 4 PRE and 4 POST neurons is first built to recognize static images of "S/J/T/U", as shown in Fig. 3b. The aim is that each POST neuron only fires a spike in responding to a specific input pattern and remains silent for the others. We employ the ReSuMe algorithm[39], a variant of bio-inspired spike-timing-dependent plasticity (STDP), to perform on-chip supervised learning. Compared with traditional gradient-based algorithms, spike-based approaches leverage the timing information to train PSNNs and offer distinct advantages of sparsity and efficiency in overall spiking dynamics. As shown in Fig. 3c, synaptic weights are depressed or strengthened based on the time differences between pre- and postsynaptic spikes, whenever the neuron fails to respond to the same desired state as the teacher. Figure 3d presents the firing states during the training process. Each POST neuron only generates random output spikes at the beginning and progressively shows a preference for a certain input pattern as the iteration epoch increases. The measured waveforms after convergence are given in Fig. 3e, where all POST neurons succeed in distinguishing a specific pattern as expected. For instance, POST1 fires a strong spike only if the pattern "S" is shown, and remains sub-threshold oscillations to the others.

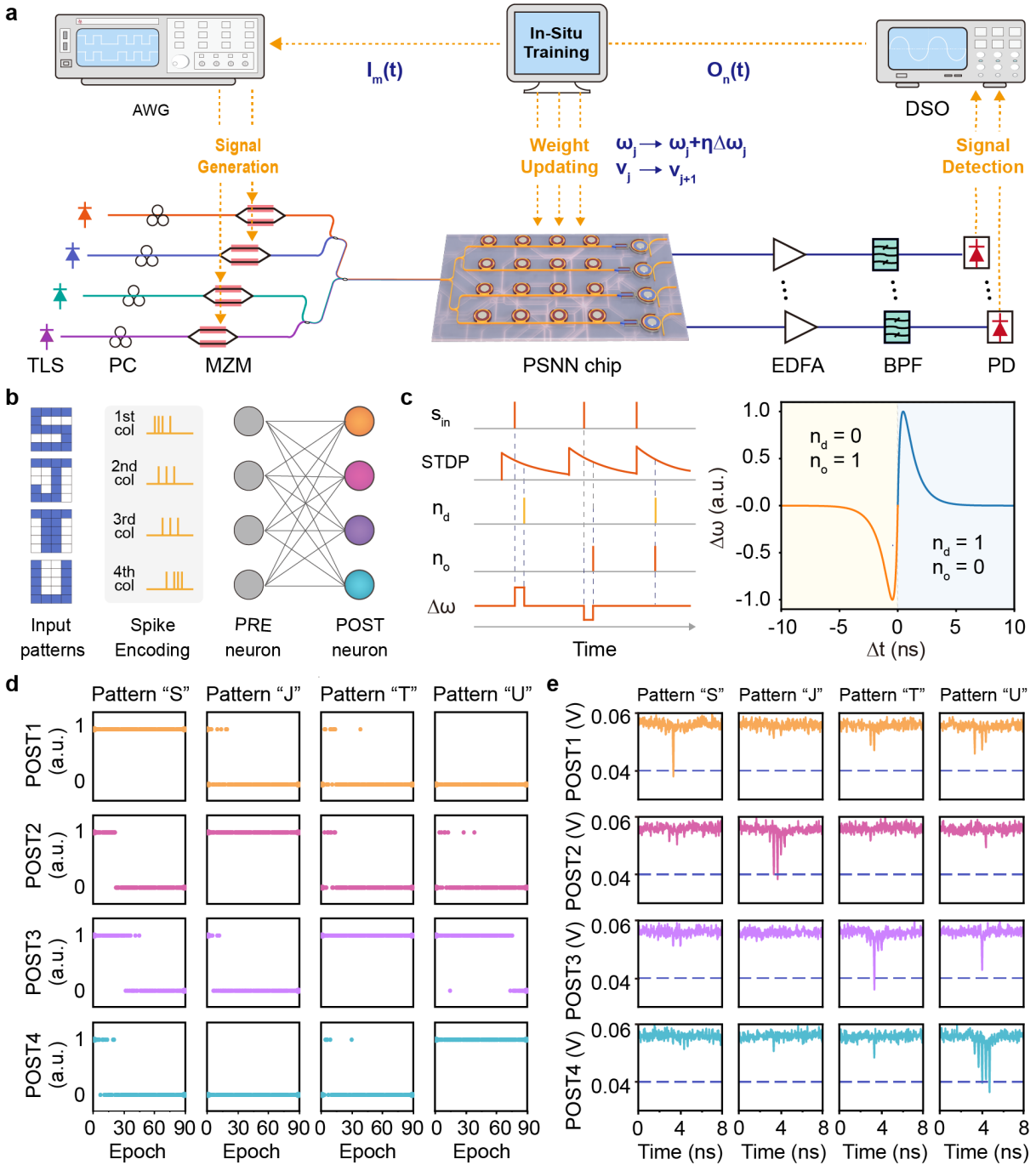

**Fig. 3 | Supervised *in-situ* training of the PSNN chip for pattern recognition. a**, Experimental setup for in-situ PSNN training, which incorporates input preparation, signal generation, on-chip photonic processing, data acquisition, and weight updating into a closed control pipeline. TLS, tunable laser source; PC, polarization controller; MZM, Mach-Zehnder intensity modulator; EDFA, erbium-doped fiber amplifier; BPF, bandpass filter; PD, photodetector; AWG, arbitrary waveform generator; DSO, digital storage oscilloscope. **b**, Schematic of the input patterns, spike encoding, and network architecture. **c**, Supervised synaptic plasticity learning rule. When the actual output state $n_o$ is not the same as the desired state $n_d$, the weight is strengthened based on the time differences between them if the pre-spike arrives before the post-spike, otherwise depressed. **d**, Activation states of the four POST neurons in responding to different input patterns during the in-situ training process. **e**, Measured output waveforms of the four POST neurons after training, where each POST neuron only generates a strong spike response to a certain input pattern.

## Human action video recognition

Ultrafast video processing has important applications in areas like surveillance, control, and behavior analysis. The PSNN chip, by exploiting a bio-inspired spike encoding technique, holds great potential to process continuous streams of dynamic vision data, much like how our brains effortlessly handle the ever-evolving stimuli of the real world. Conventional image sensors (Fig. 4a) produce image frames at fixed time intervals—whether a feature has changed or not—thus generating a large amount of redundant vision data but limited information. In contrast, ganglion cells in the retina (Fig. 4b) deliver a spike only when and where a certain feature in their receptive field changes[40], hence offering great advantages of low motion blur, high dynamic range, and spatiotemporal sparsity, all while using less bandwidth and power resources. Inspired by the unique mechanism of retinas, we propose to encode only temporal variations of video frames by registering the times and spatial locations of changes at the pixel level. The absence of spike events when no change of contrast is detected implies that redundant information recorded in original frames is not carried in the encoded stream. More importantly, this sparse and unsynchronized manner naturally aligns with event-based PSNNs and has significantly alleviated the burden of input representations.

We first record a homemade video containing five actions (i.e., handwaving, handclapping, handcrossing, boxing, and kicking) to demonstrate human action recognition (Movie S1). Figure 4c illustrates the working principle. In the pre-processing stage, we extract 40 keyframes for each action from the original video and employ an attention mechanism to generate binary silhouettes by discarding the background and horizontally centering the silhouettes in the frame (See supplementary Note 9 for more detailed information). As such, we remove all unnecessary information, such as background clutter or shadows, and focus the attention of the PSNN classifier on the pose of the subject, disregarding the influence of its spatial position in raw frames. After that, 40 binary frames are obtained for each action with a pixel size of 64 × 64. Guided by the event-driven encoding technique, we subtract the pixel intensities of each frame from the previous one and obtain 39 binary feature maps by taking the absolute values. Figure 4d presents five encoded frames for the action of "kicking", where only regions of body movements are highlighted as expected. The feature maps are flattened into a one-dimensional input vector and fed into a fully connected output layer consisting of 5 postsynaptic neurons. The final classification result is indicated by the output neuron that fires a spike for a specific action.

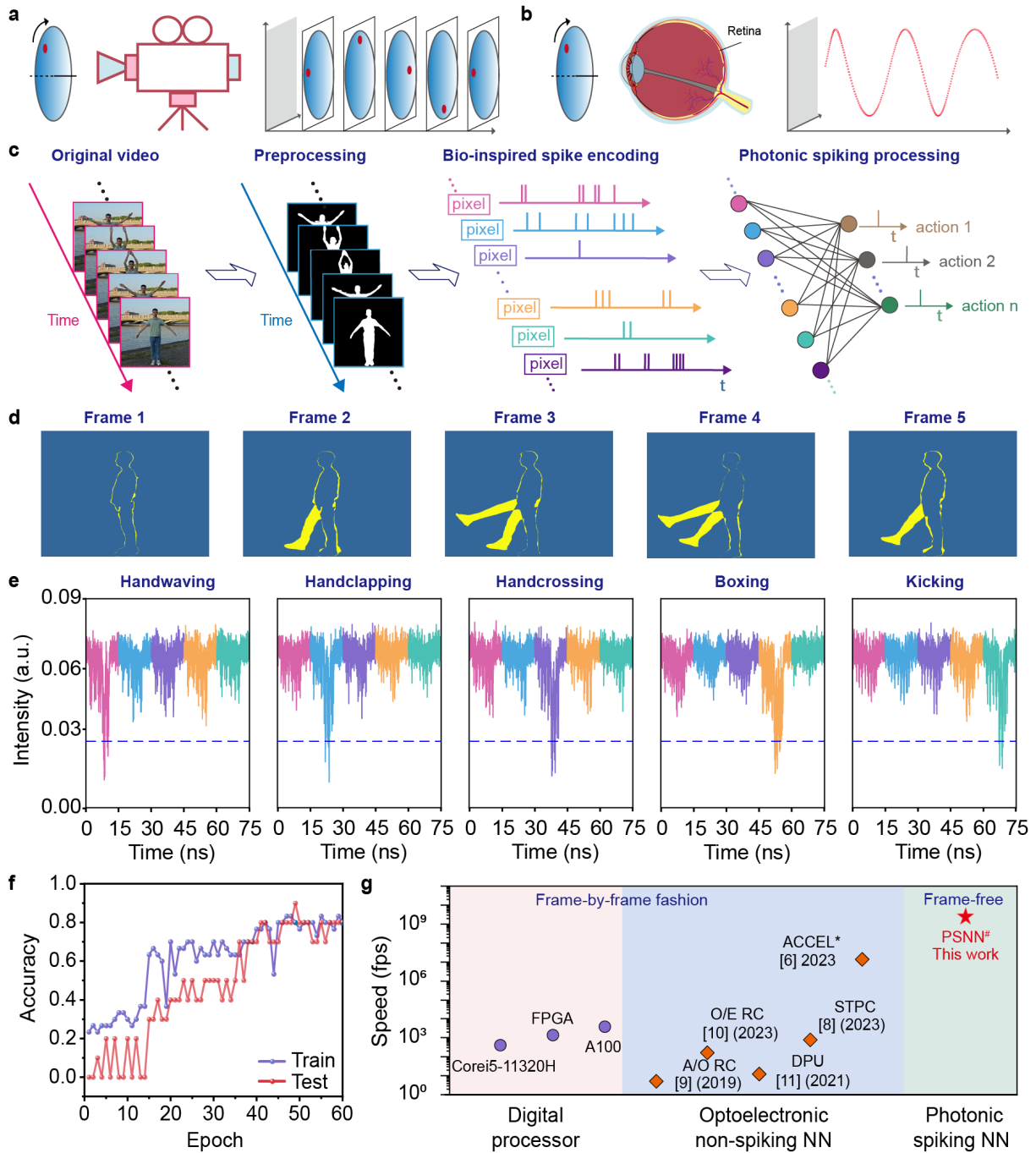

**Fig. 4 | Principles and experimental results of human action video recognition.** The working mechanisms of **a,** standard cameras and **b,** biological retinas. **c,** Workflow of the constructed PSNN. **d,** Feature maps of the "Kicking" action obtained from the retina-inspired spike encoding. **e,** Captured waveforms of 5 MRM neurons in response to different actions. **f,** Accuracy on the KTH dataset during the in-situ training process. **g.** Comparison of video processing speed on different hardware platforms.

CPU: Intel Core i5-11320H[8]; FPGA: Xilinx Virtex-6 LX760[41]; A/O RC: All-optical reservoir computer based on a spatial light modulator[9]; O/E RC: Optoelectronic reservoir computer based on a Mach-Zehnder Modulator[10]; DPU: Diffractive processing unit with five recurrent blocks[11]; STPC: spatial-multiplexed spatiotemporal photonic computing module[8]; ACCEL: all-analog chip combining electronic and light computing[6], note that a digital NN is needed for the decision making. PSNN: Photonic spiking neural network, note that the linear matrix multiplication is currently implemented in the digital domain.

Limited by the current scale of integrated photonic synapse arrays, the weight matrix size of 4096×5 is beyond the capabilities of a lab demonstration. Without sacrificing the core concept of fully bioinspired spiking information processing on a silicon photonic chip, we make a compromise by performing the linear matrix multiplication in the electrical domain and the nonlinear spike activation in the optical domain. The weighted spike trains are converted into a time-multiplexed AWG sequence and then directly applied to an MRM neuron, which successively emulates 5 postsynaptic neurons in different time slots. Note that the principle used here allows a hardware-friendly implementation of neuromorphic photonic systems and has already enabled many fantastic applications[20,37]. The acquired results after in-situ training are given in Fig. 4e, where output waveforms of five optoelectronic neurons are presented with different colors, and the threshold value of 0.026 is marked with a dotted line. Obviously, each neuron delivers a strong spike response only to specific action video frames, while generating sub-threshold oscillations to the others. Moreover, we validate the scalability of the PSNN system on the popular KTH dataset. Specifically, 40 videos shot over a uniform background are chosen from the first scenario of three motions (boxing, handwaving, and handclapping), and 30 videos are randomly selected as the training dataset, with the remaining 10 samples as the test dataset. Figure 4f presents the accuracy during the in-situ training process. After convergence, our PSNN has achieved an impressive classification accuracy of 80% on both the training and test datasets.

**Discussion**

Achieving ultrafast data processing with high energy efficiency remains the ultimate goal for neuromorphic photonic computing. The core PSNN chip can exhibit a computing density of 114.3 GMAC/s/mm$^2$, coupled with an energy efficiency exceeding 2.75 GSOP/W, attributed to advanced integration technology and well-optimized photonic devices (See Supplementary Note 12 for more details). In our experiment, the pixel value for each frame occupies a time slot of 0.28 ns, corresponding to a frame rate of 3.57 GHz. The time duration to classify a video with 40 keyframes is 0.28 ns × 39 (time steps)× 3 (neuron number) + 15 ns × 2 (zero padding) = 62.76 ns, resulting in an overall video processing speed of 15.93 MHz.

Figure 4g compares the video inference speed, i.e., FPS, between existing hardware platforms. Integrated and fiber-based photonic convolutional accelerators[27,42], as well as the all-analog optoelectronic chip[6], have demonstrated high throughput for video frame processing, but their overall speed is often hindered by frequent read-compute-write operations of memory and the need for digital post-processing in decision-making. Though free-space optoelectronic diffractive processing units[11], all-optical reservoir computer[9], and spatiotemporal photonic computing system[8] capitalize on high parallelism, they are constrained by digital electronics with slow refresh rates, such as DMDs or cameras. The inter-frame processing interval for frame-based sensors typically spans several milliseconds, which is inadequate for real-time analysis of ultrafast visual scenes. It's worth mentioning that processing videos in a sequential frame-by-frame manner also results in high information redundancy and leads to more energy consumption. In stark contrast, our PSNN chip can fully harness the potential of high-speed light modulation and propagation by processing input videos with sparse spatiotemporal events, and the frame-free processing fashion yields a dramatic speed improvement—by approximately two orders of magnitude—over prior frame-based computing paradigms.

Admittedly, the classification accuracy of a single-layer PSNN is modest, yet this can be further improved by increasing the network depth, incorporating convolutional or pooling layers, and applying more advanced training algorithms. As a few thousand photonic elements have already been demonstrated on the silicon platform[43], the proof-of-concept PSNN chip can be easily scaled up with available well-optimized device libraries and mutual fabrication technologies. It's worth mentioning that the spike encoding technique in our work is very close to the working principles of neuromorphic vision sensors that generate asynchronous spikes by detecting pixel-level brightness changes[44,45]. By integrating the silicon retina as the "eye" of our PSNN, it becomes increasingly promising to build a standalone neuromorphic processing system that is field-deployable for diverse real-time applications like visual navigation in autonomous vehicles.

In conclusion, we report the first GHz spiking neuromorphic photonic chip capable of processing dynamic visual information in a biologically plausible event-driven manner. The PSNN chip introduces ultrafast (~4 GHz), scalable, and integrated optoelectronic spiking neurons based on carrier-injection MRMs, eliminating the need for III-V or other exotic materials like PCMs. The optical-electrical-optical conversion allows on-chip regeneration of inputs to the next layer without optical amplification, inherently supporting the cascadability for multi-layer networks. Our PSNN relies entirely on CMOS-compatible foundry fabrication, which can potentially flourish into wafer-level spiking processing systems. Besides, we have developed the first supervised in-situ training scheme for PSNNs by extracting precise time information from pre- and post-synaptic spikes, mitigating noise influences in analog systems. The training approach, though solely applied to single-layer cases in our work, can be easily generalized to deep PSNNs following the same concept. Moreover, we have realized frame-free video processing by converting videos into spatiotemporal spike representations, much like retinas do efficiently, achieving an impressive 80% accuracy on the KTH dataset. Compared to existing computing approaches, the PSNN achieves a two-order-of-magnitude speed improvement, enabling real-time processing of ultrafast dynamic vision data. The PSNN chip not only bridges the gap between neuroscience and computer science, but also holds great promise for low-latency, low-power, and high-speed edge applications in modern society, marking a significant breakthrough in the development of neuromorphic optoelectronic hardware.

## Methods

### Characterization of the MRM spiking neuron

Grating couplers are employed to interface single-mode fibers and bus waveguides of MRM neurons, with an optimized coupling loss of ~7 dB/facet at 1550 nm. The pump light from a continuous-wave tunable laser source (Santec, TSL770) is directly fed into the photonic chip after being polarized by a polarization controller. While the perturbation light from another laser (Keysight, M81960A) is modulated with a high-speed MZI intensity modulator (Fujitsu, FTM7939EK, 25 GHz) and converted into a current signal by a photodetector (Finisar, XPDV2120R), which is then applied to the MRM through an RF probe. The electrical modulation signals are generated by an arbitrary waveform generator (AWG, Keysight, M8195A, 25 GHz, 65 GSa/s) and boosted by an RF amplifier (SHF, S807C, 55 GHz). The output of the MRM neuron is first connected to an erbium-doped fiber amplifier (EDFA) to compensate for optical loss, then fed into a bandpass filter (DiCon, TF-1550-0.8-9/9LT-FC-1) centered at the pump wavelength to filter out noise. The optical signal is finally detected by a photodetector (Finisar, XPDV2120R) and then sent into a digital oscilloscope (Lecroy, LabMaster 10-Zi-A, 36 GHz, 80 GSa/s) to capture real-time waveforms. Two LAN cables are used to form communication links between AWG, the oscilloscope, and the host desktop computer (Windows 10, Intel Core i7-8700). The MRM is pumped with a proper wavelength and power to work in its excitable regime, and perturbation pulses with varying strengths and time durations are applied to the MRM to investigate its neuron-like spiking dynamics.

### Characterization of the MRR synapse array

The ten resonance wavelengths of a silicon MRR weight bank are first thermal tuned to realize almost equal channel spacing by applying proper voltages. To obtain the "weight-voltage" lookup table of each channel, the pump wavelength is set to be the working wavelength, i.e., 1550.170 nm for the first channel and 1557.370 nm for the last channel, and the applied voltage on the corresponding MRR is swept from 0 V to 3.6 V with a step of 0.01 V, with the other nine voltages remaining unchanged. For a given weight matrix, the initial voltage of each MRR is determined by its "weight-voltage" lookup table, and an in-situ calibration method (More details are provided in Supplementary Note 2) is used for fine-tuning to compensate inter-channel thermal crosstalk.

### Pattern recognition task

The experimental setup is shown in Fig. 3a, where the working wavelengths of a four-channel tunable laser (IDPhotonics, DX) are set to be 1550.170 nm, 1550.970 nm, 1551.770 nm, and 1552.570 nm, respectively. In the temporal encoding process, each column of the $5 \times 4$ binary pixel matrix is converted into a spike train and fed into a PRE neuron. For better clarity, the encoded spatiotemporal event sequences of pattern "S" are also given in Fig. 3b. The encoded four spike trains are modulated onto four optical carriers via intensity modulators (Fujitsu, FTM7939EK, 25 GHz). Specifically, each black pixel is converted into a rectangular pulse of 0.1 ns, and the center time is determined by $t = t_s \cdot (x + y) \, ns$, where $x$ and $y$ are the corresponding subscript index in the pixel matrix, respectively. The term $t_s$ denotes the time duration between two neighboring pixel representations and is set to be 0.35 ns in our experiment. To determine the start time point of the acquired waveform for each training epoch, a padding sequence consisting of 3 rectangular pulses of 0.4 ns is added before the encoded pattern. Constrained by available instruments in our lab, each time only one row of the 4×4 weight matrix is trained. The multiplexed signal is weighted by four channels of the MRR weight bank and then converted into a current signal by a photodetector (Finisar, XPDV2120R). The generated photocurrent is applied to the MRM neuron, and the corresponding output waveform is measured one by one. Besides, tunable attenuators and delay lines are used to balance optical powers and time delays between 4 signal paths, which are not included in Fig. 3a.

**Human action video recognition**

The experimental setup is the same as that for MRM neuron characterization. The weighted-and-summed spike trains of input videos are converted into a time-multiplexed AWG sequence and then directly applied to an MRM neuron, which successively emulates all post-synaptic neurons in different time slots. Note that the principle used here allows a hardware-friendly implementation of neuromorphic photonic systems and has enabled many fantastic applications[20,37]. Besides, a change of pixel value between two frames is represented by a rectangular pulse of 0.15 ns within a time window of 0.28 ns. Therefore, each pixel is encoded into a 10.92 ns temporal spike train of 39 time steps.

**Modified ReSuMe algorithm for weight updating**

During the PSNN training process, the synapse weights are updated according to the time differences of spike response between pre- and postsynaptic neurons. In each learning epoch, the change in synaptic efficiency $\omega_i$ can be expressed as follows:

$$\Delta\omega_i = \begin{cases} \omega_f \cdot \sum_{t_i \leq t_{min}} K(t_{min} - t_i), & if\ n_d = 1\ and\ n_o = 0 \\ -\omega_f \cdot \sum_{t_i \leq t_{min}} K(t_{out} - t_i), & if\ n_d = 0\ and\ n_o = 1 \end{cases}$$

where $n_d$ and $n_o$ denote the desired and actual number of output spikes, respectively. Here $t_i$ is the presynaptic spike time, $t_{out}$ represents the actual output spike time, and $t_{min}$ denotes the time at which the neuron output reaches its minimum value. The learning rate $\omega_f$ specifies the maximum amount of the synaptic update and is set to be $0.4 \times 10^8$. The learning window determined by the K function only considers spikes $t_i \leq t$, and is described as:

$$K(t) = V_0 \cdot \left(\exp\left(-\frac{t}{\tau_m}\right) - \exp\left(-\frac{t}{\tau_s}\right)\right),$$

where $V_0 = 2.1165$, $\tau_m = 1\ ns$, $\tau_s = 0.25$ ns. In cases where the POST neuron is desired to fire a spike ($n_d = 1$), the corresponding synaptic weight will be increased by $\Delta\omega_i$ if no output spike is observed ($n_o = 0$). Conversely, the weight will be decreased by $\Delta\omega_i$ if an output spike appears ($n_o = 1$) when the POST neuron is expected to keep silent ($n_d = 0$).

**Acknowledgments**

This work was supported by the National Research and Development Program of China (2023YFB2804702); National Natural Science Foundation of China (NSFC) (62175151, 62341508，62422509); Shanghai Municipal Science and Technology Major Project; Shanghai Frontiers Science Center Program(2021–2025 No. 20) .

**Author contributions**

X.H.G., M.G., and Y.K.S initiated and supervised the project. X.H.G. and J.L.X. conceived the research. J.L.X. designed and characterized the silicon chip. J.L.X., J.X., Y.L.C., A.H., and Y.T.Z. conducted the experiments. J.L.X., J.X., and Y.L.C. processed the data. J.L.X., X.Y.F., M.G., Y.K.S, and X.H.G. discussed the results. All authors contributed to writing the manuscript.

**Competing interests**

The authors declare no competing interests.

**Data Availability**

The KTH dataset is publicly available and can be accessed at https://www.csc.kth.se/cvap/actions/.s. Data underlying the results presented in this paper may be obtained from the authors upon reasonable request.

**Additional information**

**Supplementary information** The online version contains supplementary material is available.
**Correspondence and requests for materials** should be addressed to Xuhan Guo.